\documentclass[twocolumn,twocolappendix]{aastex7}
\submitjournal{ApJ}

\begin{document}

\title{The Aromatic Infrared Bands around the Wolf-Rayet Binary WR\,140 Revealed by JWST}

\author[0000-0003-4402-6475]{Kotomi Taniguchi}
\affiliation{National Astronomical Observatory of Japan, National Institutes of Natural Sciences, 2-21-1, Osawa, Mitaka, Tokyo, Japan}
\email[show]{kotomi.taniguchi@nao.ac.jp}

\author[0000-0002-2106-0403]{Ryan M. Lau}
\affiliation{NSF's NOIRLab, 950 N Cherry Ave, Tucson, 85719, Arizona, USA}
\affiliation{IPAC, Mail Code 100-22, Caltech, 1200 E. California Blvd., Pasadena, CA 91125, USA}
\email[show]{ryanlau@ipac.caltech.edu}

\author[0000-0002-8234-6747]{Takashi Onaka}
\affiliation{Department of Astronomy, Graduate School of Science, University of Tokyo, Tokyo, Japan}
\email{onaka@astron.s.u-tokyo.ac.jp}

\author[0000-0003-4801-0489]{Macarena Garcia Marin}
\affiliation{European Space Agency, Space Telescope Science Institute, Baltimore, MD 21218, USA}
\email{maca@stsci.edu}

\author[0009-0004-1611-4229]{Hideo Matsuhara}
\affiliation{Institute of Space and Astronautical Science, Japan Aerospace Exploration Agency, 3-1-1 Yoshinodai, Chuo-ku, Sagamihara, Kanagawa 252-5210, Japan}
\email{maruma@ir.isas.jaxa.jp}

\author[0000-0002-4333-9755]{Anthony Moffat}
\affiliation{Centre de Recherche en Astrophysique du Québec, Département de physique, Université de Montréal, Complexe des Sciences, Montréal, QC H2V 0B3, Canada}
\email{Moffat@umontreal.ca}

\author{Theodore R. Gull}
\affiliation{NASA Goddard Space Flight Center, Greenbelt, MD 20771, USA}
\email{tedgull@gmail.com}

\author[0000-0001-7697-2955]{Thomas I. Madura}
\affiliation{Department of Physics and Astronomy, San Jos\'e State University, San Jos\'e, CA 95192-0106, USA}
\email{thomas.madura@sjsu.edu}

\author[0000-0001-9754-2233]{Gerd Weigelt}
\affiliation{Max Planck Institute for Radio Astronomy, Auf dem H\"ugel 69, 53121 Bonn, Germany}
\email{weigelt@mpifr.de}

\author{Riko Senoo}
\affiliation{Institute of Astronomy, Graduate School of Science, University of Tokyo, 2-21-1 Osawa, Mitaka, Tokyo 181-0015, Japan}
\email{senoo-riko222@g.ecc.u-tokyo.ac.jp}

\author[0000-0001-8136-9704]{Alan T. Tokunaga}
\affiliation{Institute for Astronomy, University of Hawaii, 2680 Woodlawn Dr., Honolulu, HI 96822}
\email{tokunagaa001@gmail.com}

\author{Walter Duley}
\affiliation{Department of Physics and Astronomy, University of Waterloo, 200 University Avenue West, Waterloo, Ontario, Canada N2L3G1}
\email{wwduley@uwaterloo.ca}

\author[0000-0002-8092-980X]{Peredur M. Williams}
\affiliation{Institute for Astronomy, University of Edinburgh, Royal Observatory, Edinburgh, EH9 3HJ, UK}
\email{pmw@roe.ac.uk}

\author[0000-0002-2806-9339]{Noel D. Richardson}
\affiliation{Department of Physics and Astronomy, Embry-Riddle Aeronautical University, 3700 Willow Creek Rd, 
Prescott, AZ 86301, USA}
\email{noel.richardson@erau.edu}

\author[0000-0002-9723-0421]{Joel Sanchez-Bermudez}
\affiliation{Universidad Nacional Aut\'onoma de M\'exico. Instituto de Astronom\'ia. A.P. 70-264, 04510, Ciudad de M\'exico, 04510, M\'exico}
\email{joelsb@astro.unam.mx}



\begin{abstract}
We have analyzed the aromatic infrared bands (AIBs) in the $6-11.2$ \micron\, range around the Wolf-Rayet binary WR\,140 ($d=1.64$ kpc) obtained with the James Webb Space Telescope (JWST) Mid-Infrared Instrument (MIRI) Medium-Resolution Spectrometer (MRS).
In WR\,140's circumstellar environment, we have detected AIBs at 6 \micron \, and 7.7 \micron \, which are attributed to C-C stretching modes.
These features have been detected in the innermost dust shell (Shell\,1; $\sim2100$ au from WR\,140), the subsequent dust shell (Shell\,2; $\sim5200$ au), and ``off-shell'' regions in the MRS coverage.
The 11.2 \micron\ AIB, which is associated with the C-H out-of-plane bending mode, has been tentatively detected in Shell\,2 and the surrounding off-shell positions around Shell\,2.
We compared the AIB features from WR140 to spectra of established AIB feature classes A, B, C, and D.
The detected features around WR\,140 do not agree with these established classes.
The peak wavelengths and full width half maxima (FWHMs) of the 6 \micron\ and 7.7 \micron\ features are, however, consistent with those of R Coronae Borealis (RCB) stars with hydrogen-poor conditions.
We discuss a possible structure of carbonaceous compounds and environments where they form around WR\,140.
It is proposed that hydrogen-poor carbonaceous compounds initially originate from the carbon-rich WR wind, and the hydrogen-rich stellar wind from the companion O star may provide hydrogen to these carbonaceous compounds.
\end{abstract}



\section{Introduction} \label{sec:intro}

Carbon, the fourth most abundant element in the local universe, takes various forms in the interstellar medium (ISM): ionic and atomic carbon (C$^+$, C), unsaturated carbon-chain molecules \citep{2024Ap&SS.369...34T}, polycyclic aromatic hydrocarbons \citep[PAHs;][]{2008ARA&A..46..289T}, fullerenes \citep[C$_{60}$, C$_{70}$, and C$_{60}$$^+$;][]{2010Sci...329.1180C, 2020JMoSp.36711243L}, mixed aromatic/aliphatic organic nanoparticles (MAON) and hydrogenated amorphous carbon \citep[HAC;][]{2022Ap&SS.367...16K}, and carbonaceous dust grains \citep{1981MNRAS.196..269D, 2003ARA&A..41..241D}.
Although observational studies on each species have been conducted so far, their relationships among the different forms in the ISM are unclear.
For instance, the detection of aromatic species from the molecular cloud cyanopolyyne peak in the Taurus Molecular Cloud-1 (TMC-1 CP) suggests the inheritance of PAHs from the diffuse ISM to the molecular clouds \citep[$e.g.,$][]{2021Sci...371.1265M, 2024Sci...386..810W, 2025ApJ...984L..36W}.
Since their abundances cannot be reproduced by typical chemical network simulations considering only the bottom-up molecular formation, we need the top-down molecular formation scenario.
Hence, the PAH chemistry in the ISM becomes more important in understanding the initial chemical compositions in star-forming regions \citep{Taniguchi2025}.

PAHs are a key component of the ISM. 
They typically contain 10\% of the carbon in the diffuse ISM \citep{2021A&A...650A.193O}.
Their emission features can be observed in the near- and mid-infrared regimes and are called Aromatic Infrared Bands (AIBs); the major bands are seen at 3.3, 6.2\footnote[1]{The typical aromatic C-C stretching modes fall between 6.1 and 6.5 \micron\, and we call it ``6 \micron\, feature'' in this paper.}, 7.7, 8.6, and 11.2, and 12.7 \micron\, together with a number of minor bands \citep{2008ARA&A..46..289T, 2024A&A...685A..75C}.
The spectral profiles in the 6 -- 9 and 11 -- 14 \micron\, ranges have been used for the classification; classes A, B, C \citep{2002A&A...390.1089P, 2004ApJ...611..928V}, and D \citep{2014MNRAS.439.1472M}.
These different classes have been considered to reflect different structures of complex carbonaceous compounds, and are likely related to astronomical object types:
Class A spectra are generally seen in the ISM of our Galaxy and nearby galaxies. They are also seen in post-asymptotic giant branch (AGB) stars and in OB stars associated with \ion{H}{2} regions;
class B spectra originate from carbon-rich planetary nebulae (PNe) and the disks around young stellar objects (YSOs); class C spectra arise from a variety of objects including Herbig AeBe stars, T Tauri stars, carbon-rich AGB objects, and carbon-rich red giants; class D spectra can be observed in carbon-rich post-AGB objects \citep{2022A&A...665A.153J}.
Structures of complex carbonaceous compounds change by processing in the ISM \citep[$i.e.,$ photons, energetic particles, and stellar winds;][]{2017A&A...602A..46J}.

In this paper, we focus on AIBs around the carbon-rich dust-forming Wolf-Rayet (WR) star WR\,140, which is also known as HD 193793 \citep[$d=1.64$ kpc;][]{1990MNRAS.243..662W, 2021MNRAS.504.5221T}.
WR stars have masses of 10 -- 25 M$_{\odot}$ and descended from O-type stars \citep{2007ARA&A..45..177C}.
Some of the carbon-rich WR stars, $i.e.,$ WC type stars, form carbonaceous dust by the wind-wind interaction in the binary systems with hydrogen-poor conditions \citep{1990MNRAS.243..662W, 2007ARA&A..45..177C}.
They are considered to be one of the important carbon suppliers to the ISM \citep{2020ApJ...898...74L}.
Since their lifetimes are short \citep[the WR phase lasts $\sim$ 0.5 Myr;][]{2007ARA&A..45..177C}, these dust-forming WC stars may provide carbonaceous dust in the ISM in the early universe \citep{2022NatAs...6.1308L, 2023ApJ...951...89L}.
Recent observations with the James Webb Space Telescope (JWST) toward four WC stars have shown that dust formed by these stars is long-lived, more than 300 years in some systems \citep{2025ApJ...987..160R}.
Complex carbonaceous compounds, which are potential precursors of PAHs, could also be produced by these stars.

The WR\,140 binary system consists of a carbon-rich evolved massive star with a subtype of WC7pd and an O5.5fc star \citep{2011MNRAS.418....2F}. 
Its orbit is highly eccentric ($e=0.89$) with a 7.93 yr (2895 days) orbital period, and the wind-wind collisions cause periodic dust formation \citep{2011ApJ...742L...1M, 2022NatAs...6.1308L}.
\citet{2023ApJ...951...89L} investigated the circumstellar dust properties using the Subaru telescope data.
They found that hot ($T_d \approx 1000$ K) and cooler ($T_d \approx 500$ K) components coexist, and the hot component is attributed to the nano-sized grains formed by grain-grain collisions or rotational disruption of the larger grain size population by radiative torques in the strong radiation field \citep{2019NatAs...3..766H}.  
This source was observed using the JWST, and high angular-resolution infrared data were obtained.
\citet{2022NatAs...6.1308L} presented 17 dust shells around WR\,140 with the JWST Mid-Infrared Instrument (MIRI) Medium-Resolution Spectrometer and Imager.
Their results show that carbonaceous dust grains produced by WR\,140 can survive at least $\sim130$ yr.
More recently, \citet{2025ApJ...979L...3L} derived the dust shell proper motion ($390 \pm 29$ mas yr$^{-1}$, corresponding to $2714 \pm 188$ km\,s$^{-1}$) by comparing the data obtained in JWST cycles 1 and 2.

This paper is structured as follows:
Section \ref{subsec:obs} describes the JWST/MIRI observations toward WR\,140 along with the pipeline processing and data reduction procedure.
In Section \ref{sec:res}, spectra at 11 selected positions and results of spectral analyses are presented.
We compare the spectra around WR\,140 to typical spectra of classes A--D (Section \ref{subsec:class}) and compare spectral features among different positions around WR\,140 (Sections \ref{subsec:spatial} and \ref{subsec:correlation}).
The spectral profiles of the 6 \micron \, and 7.7 \micron \, features around WR\,140 are compared with the other types of astronomical objects in Section \ref{subsec:62feature}.
We discuss the chemistry of carbonaceous compounds around WR\,140 in Section \ref{subsec:PAHWR140}.
The main conclusions of this paper are summarized in Section \ref{sec:con}.

\section{Observations \label{subsec:obs}}

MIRI Medium-Resolution Spectrometer \citep[MRS;][]{2015PASP..127..646W} observations of WR\,140 were performed on 2022 July 8 UT in JWST Cycle 1 (PID 1349)\footnote{The data is available at MAST:\dataset[doi:10.17909/gf7h-0264]{http://dx.doi.org/10.17909/gf7h-0264}.}.
All four channels (Ch 1 -- 4) and all three grating settings (SHORT, MEDIUM, LONG) were used and resulted in spatially resolved spectra from 4.9--28.1 \micron\, with a spectral resolving power of $R\approx1500-3500$. The MRS observations used a 4-point dither optimized for an extended source. An exposure was taken in each dither position using the FASTR1 readout pattern with 60 groups and 5 integrations, which corresponds to a 0.94 hr exposure time in each grating setting. The spatial pixel scale of the MRS observations ranged from 0.196--0.273\arcsec and the field of view ranged from  3.2\arcsec $\times$ 3.7\arcsec to 6.6\arcsec $\times$ 7.7\arcsec \citep{Argyriou2023}.

Two sets of MRS observations with identical exposure settings were conducted to cover the dust shells that WR\,140 formed in $\sim2016$ (Shell\,1) and $\sim2008$ (Shell\,2). The observation that covered WR\,140 and Shell\,1
was centered on the coordinates (J2000) 20$^{\rm {h}}$20$^{\rm {m}}$27\fs985, +43\degr51\arcmin15\farcs90; the observation that covered Shell\,2
was centered on 20$^{\rm {h}}$20$^{\rm {m}}$27\fs781, +43\degr51\arcmin13\farcs05. Each field set is called Offset\,1 and Offset\,2, respectively. Given the large proper motion of the dust shells ($\sim390$ mas yr$^{-1}$; \citealt{2025ApJ...979L...3L}), the two MRS observations were linked in a non-interruptible sequence.

The MRS raw data were obtained from the Mikulski Archive for Space Telescopes (MAST) and reduced with version 1.16.0 of the JWST science calibration pipeline \citep{Bushouse2023} with the 1298 CRDS context. The pipeline was run in all data using the default parameters, and switching on the cosmic rays shower detection on the Stage-1 pipeline data reduction,  and the two-dimensional residual fringe correction and pixel replacement steps were applied in Stage 2 of the pipeline. 
The final pipeline products were 12 drizzled datacubes \citep{Law2023}, one per band, that combined the two individual pointings.

\begin{figure*}[ht!]
\centering
\includegraphics[bb = 10 10 620 395, scale = 0.85]{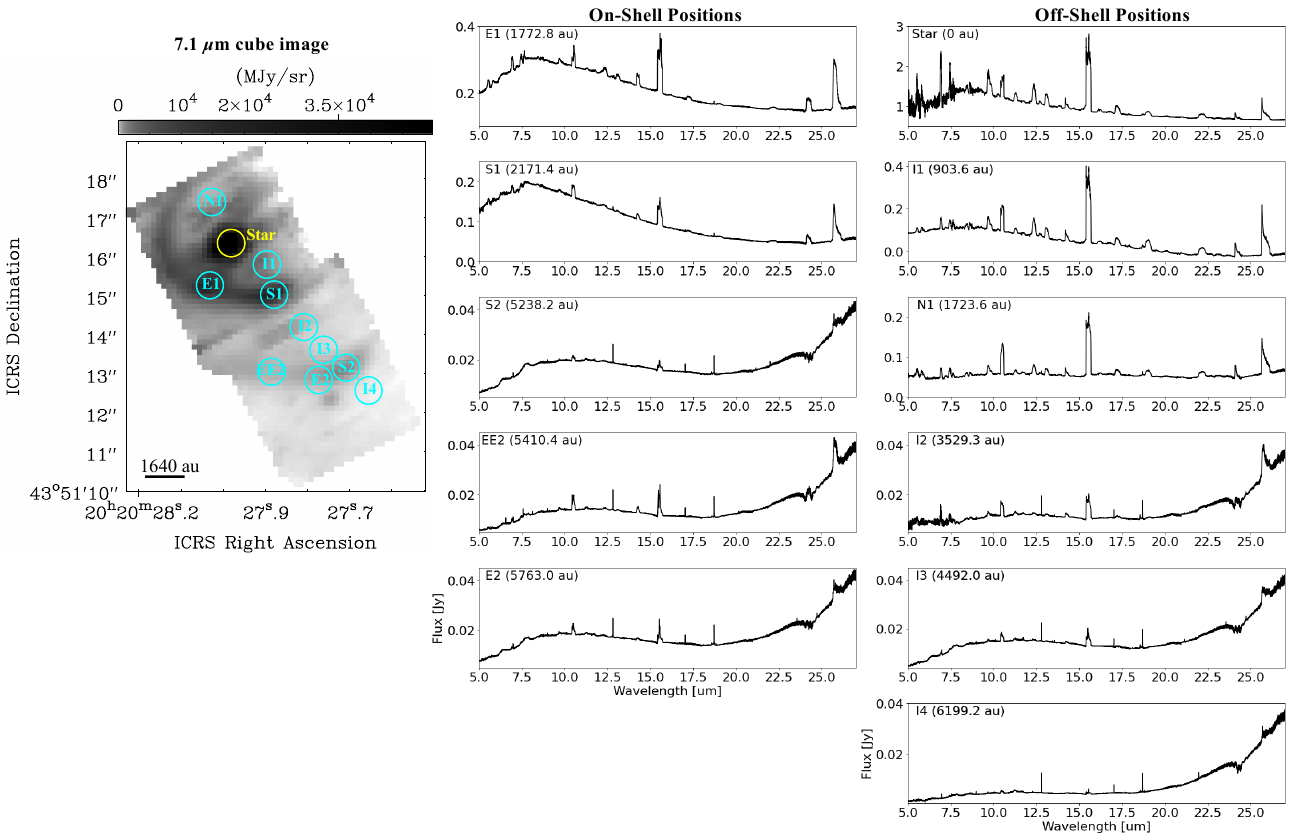}
\caption{(Left) The 7.1 $\micron$ cube image. The yellow circle indicates the position of WR\,140, and the cyan circles represent positions where we analyzed spectra. Table \ref{tab:positions} summarizes their coordinates and distances from the central binary. The radius of the circular apertures used to extract the spectra are 0.35\arcsec. (Middle and Right) The MIRI MRS spectra (5 -- 27 $\micron$) toward on-shell positions located on the dust shells of Shell\,1 and Shell\,2, and off-shell positions, respectively. The distance from the star position is indicated in each panel.
\label{fig:all}}
\end{figure*}

A circular aperture with a radius of 0.35\arcsec ($\approx575$ au) was used to extract spectra at various positions in the MRS observations of WR\,140. In order to prevent 
potential over-subtraction of possible extended circumstellar dust emission and/or AIB features within the field of view, no background subtraction is performed on the extracted spectra.
Table \ref{tab:positions} summarizes the positions where we extracted the spectra. 
We calculated projected distances of each position from the central stars in astronomical units, assuming that the distance of WR\,140 is 1.64 kpc \citep{2021MNRAS.504.5221T}.
These positions are indicated in the left panel of Figure \ref{fig:all}, which shows the 7.1\,\micron\, image of the dust continuum emission.
The S1 and S2 positions correspond to the 7.1 \micron\, emission peaks on Shell\,1 (the innermost dust shell) and Shell\,2 (the second inner dust shell), respectively.
The E1 position is to the east of S1 along Shell\,1, whereas E2 and EE2 are east of S2 along Shell\,2.

The spectra extracted from the 3 sub-bands from each of the four channels were combined additively based on the median flux density in the overlapping wavelengths between the 12 sub-bands. Extracted spectra were scaled to the Channel 1 LONG sub-band which covers the continuum between the 6.3 and 7.7 $\mu$m AIB features ($\lambda=6.53-7.65$ $\mu$m).

\begin{deluxetable}{cccccc}
\tablewidth{0pt} 
\tablecaption{Coordinates of regions for spectral analysis \label{tab:positions}}
\tablehead{
\colhead{Position} & \colhead{R.A. (J2000)} & \colhead{Decl. (J2000)} & \colhead{Distance (\arcsec)} & \colhead{$D$ (au)\tablenotemark{a}} & \colhead{Position\tablenotemark{b}} 
}
\startdata 
Star & 20$^{\rm {h}}$20$^{\rm {m}}$27\fs983 & +43\degr51\arcmin16\farcs333 & 0 & 0 & Off \\
I1 & 20$^{\rm {h}}$20$^{\rm {m}}$27\fs900 & +43\degr51\arcmin15\farcs782 & 0.55 & 903.6 & Off \\
N1 & 20$^{\rm {h}}$20$^{\rm {m}}$28\fs029 & +43\degr51\arcmin17\farcs384 & 1.05 & 1723.6 & Off \\
E1 & 20$^{\rm {h}}$20$^{\rm {m}}$28\fs035 & +43\degr51\arcmin15\farcs252 & 1.08 & 1772.8 & On \\
S1 & 20$^{\rm {h}}$20$^{\rm {m}}$27\fs883 & +43\degr51\arcmin15\farcs009 & 1.32 & 2171.4 & On \\
I2 & 20$^{\rm {h}}$20$^{\rm {m}}$27\fs811 & +43\degr51\arcmin14\farcs181 & 2.15 & 3529.3 & Off\\
I3 & 20$^{\rm {h}}$20$^{\rm {m}}$27\fs766 & +43\degr51\arcmin13\farcs594 & 2.74 & 4492.0 & Off \\
S2 & 20$^{\rm {h}}$20$^{\rm {m}}$27\fs707 & +43\degr51\arcmin13\farcs139 & 3.19 & 5238.2 & On \\
EE2 & 20$^{\rm {h}}$20$^{\rm {m}}$27\fs887 & +43\degr51\arcmin13\farcs034 & 3.30 & 5410.4 & On\\ 
E2 & 20$^{\rm {h}}$20$^{\rm {m}}$27\fs777 & +43\degr51\arcmin12\farcs819 & 3.51 & 5763.0 & On \\
I4 & 20$^{\rm {h}}$20$^{\rm {m}}$27\fs659 & +43\degr51\arcmin12\farcs553 & 3.78 & 6199.2 & Off \\
\enddata
\tablenotetext{a}{The values are calculated applying the distance to WR\,140 of 1.64 kpc \citep{2021MNRAS.504.5221T}. These values correspond to the projected distances.}
\tablenotetext{b}{``On'' means positions on the dust shells (Shell\,1 or Shell\,2), whereas ``Off'' means positions which are not on the dust shells.}
\end{deluxetable}

\section{Results and Analyses \label{sec:res}}

\subsection{Continuum Subtraction and Gaussian Fitting \label{subsec:result}}

The extracted spectra at all of the positions are presented in Figure \ref{fig:all}.
We divided the positions into two groups: on-shell and off-shell positions.
The on-shell positions include the positions on either Shell\,1 (S1 and E1) or Shell\,2 (S2, E2, and EE2).
The other positions are off-shell positions.

The broad dust emission feature peaking at $\sim7.5$ \micron\, is prominent around Shell\,1.
Although the projected distances of N1 and E1 from the central stars are almost the same, only E1 shows continuum emission from dust.
This difference between N1 and E1 is likely due to non-uniform dust formation around WR\,140 \citep{2022Natur.610..269H}.

In order to model the continuum, we fitted a spline function to each spectrum for a narrow spectral span, $i.e.,$ 3.5--4 \micron.
The results of the continuum fitting are shown in Figure \ref{fig:contfit} in Appendix \ref{appA}.
Note that thermal emission and/or scattering from the telescope system of the JWST \citep{Rigby2023} dominates the measured continuum at wavelengths $\lambda\gtrsim20$ $\mu$m in regions of the fainter areas around Shell\,2 (I2, I3, S2, E2, EE2, and I4; see Figure \ref{fig:all}). Therefore, the wavelength ranges $\lambda\gtrsim20$ $\mu$m are not used for the analysis in this work. There are also no prominent known AIBs in that wavelength range.

\begin{figure*}[ht!]
\includegraphics[bb = 10 10 785 280, width=\textwidth]{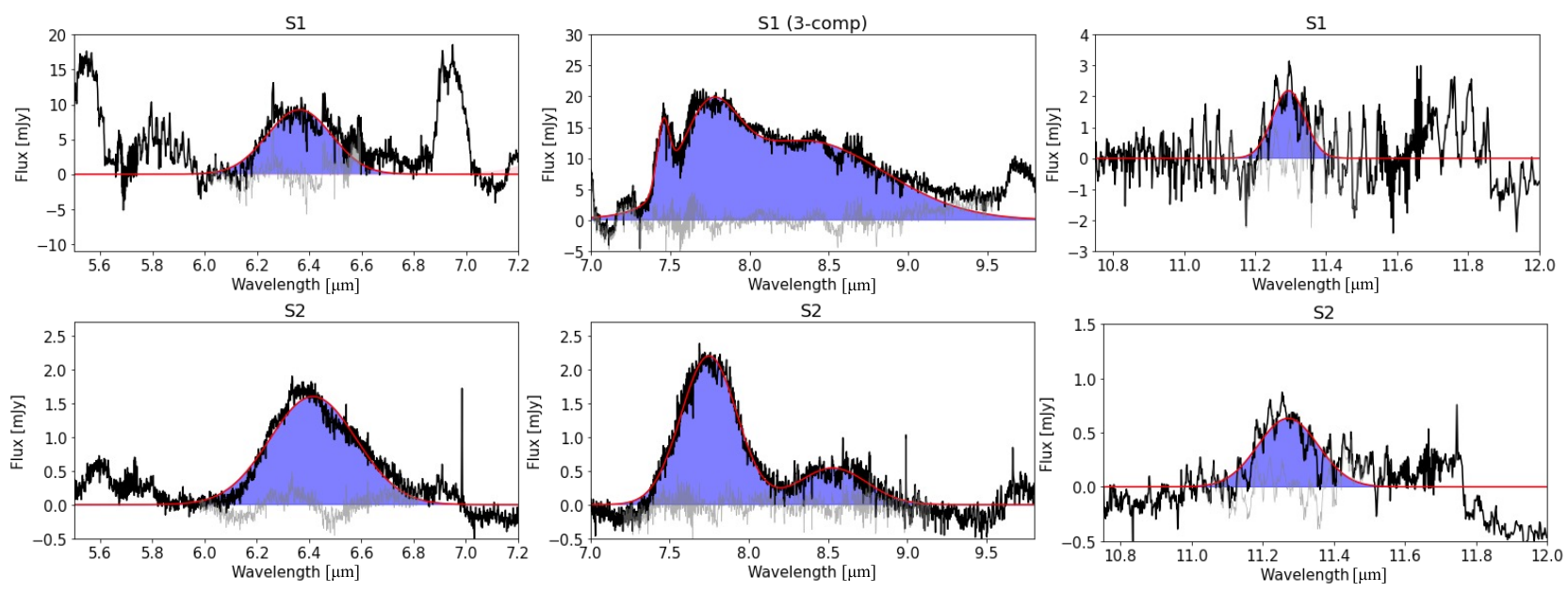}
\caption{Comparison of AIBs at S1 (on Shell\,1) and S2 (on Shell\,2). Black lines and the red curves indicate the observed spectra and the Gaussian fitting results, respectively. The filled regions highlight the Gaussian fitting results. The gray lines indicate residuals.
\label{fig:S1S2}}
\end{figure*}

We fit apparent features using single- or multiple-component Gaussian functions. 
Since we focus on the AIBs, we fitted features seen at wavelengths around the typical AIBs (6, 7.7, 8.6, and 11.2 \micron).
Table \ref{tab:gaus} in Appendix \ref{appA} summarizes line parameters obtained with the Gaussian fitting.
Figures \ref{fig:Gauss1} and \ref{fig:Gauss2} show the fitting results at all of the positions (Appendix \ref{appA}). 
We show spectra at S1 and S2, which are located at the regions of peak 7.1 \micron\, emission in Shell\,1 and Shell\,2, respectively, as examples in Figure \ref{fig:S1S2}.
Red curves and filled regions show the Gaussian fitting results.
We conducted the continuum subtraction separately in each panel.
Around the 7.5--8.6 \micron\, feature at S1, we applied a 3-component Gaussian fitting (the upper middle panel).
For the spectra at S2, a 2-component Gaussian fitting was applied.

\subsection{Properties of detected AIBs \label{subsec:identify}}

\begin{figure*}[ht!]
\includegraphics[bb = 10 5 560 390, width=\textwidth]{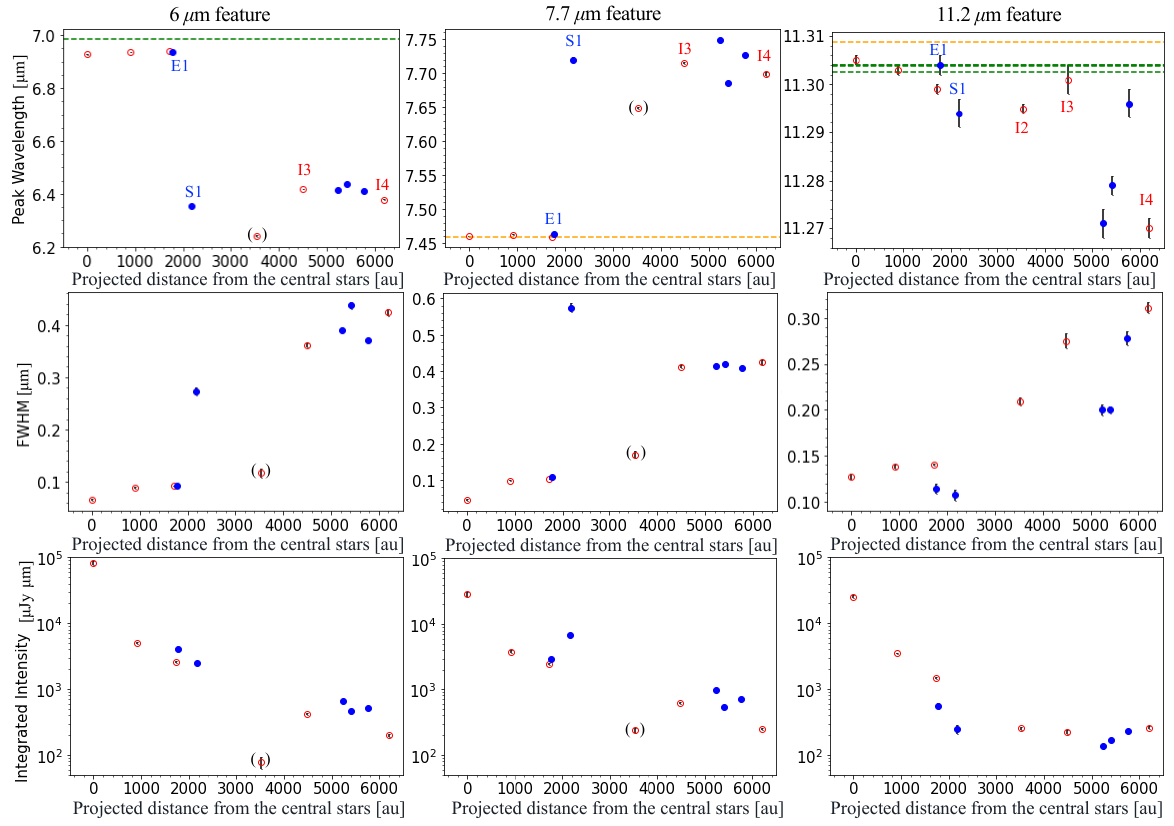}
\caption{Plots of projected distance from the central stars vs. peak wavelength (upper), FWHM (middle), and integrated intensity (lower) obtained from the Gaussian fitting for three features. The blue-filled circle and red open circle indicate positions on-shell and off-shell positions, respectively. The green dashed horizontal line in the top left panel indicates the peak wavelengths of He\,\textsc{ii} = 6.95 \micron. The orange dashed line in the top middle panel shows the wavelength of He\,\textsc{ii} = 7.457 \micron, and that in the top right panel indicates He\,\textsc{ii} = 11.304 \micron, respectively. The green dashed lines in the top right panel indicate He\,\textsc{ii} and He\,\textsc{i} lines. Since the I2 position does not show a clear emission feature at 6 and 7.7 \micron\ and the fitting results are poor, we indicate parentheses. All of the plots include error bars, but most of them are smaller than the data points. The E1 position is contaminated by atomic emission lines.
\label{fig:spefeature}}
\end{figure*}

We plotted the peak wavelength, line width (Full Width at Half Maximum; FWHM), and integrated intensity obtained from the Gaussian fitting to study the observed features in terms of the AIBs.
It should be noted that WR 140 has a large terminal wind velocity ($\sim 2800$\,km\,s$^{-1}$), which broadens atomic emission lines to a width that makes it difficult to distinguish them from relatively narrow dust features.
Therefore, we need to investigate their widths and dependence on the distance from the star carefully.
In the case of the 8.6 \micron \, feature, the double-peak features are recognized in all of the spectra, even around the star positions. 
The 8.6\,\micron \, AIB in the ISM does not show a double peak.  The apparent double peak in the present spectra could be an artifact in the data reduction process.  Therefore, we exclude the 8.6 \micron \, feature in the following discussion.

Figure \ref{fig:spefeature} shows fitting results of the of the three emission features (6, 7.7, and 11.2 \micron).
The spectrum around 5 -- 8 \micron\, at the I2 position was affected by the spectral fringing artifact \citep{Argyriou2023}, and the 6 \micron\, and 7.7 \micron\, AIBs are not clearly seen (Figure \ref{fig:Gauss2}).
Therefore, we show the I2 data with brackets. 
The top panels show the peak wavelengths of each AIB.
We show the wavelengths of the atomic lines that are close to each AIB as the horizontal lines in each panel.

We can see clear differences between positions close to the star (Star, N1, E1, and I1) and those farther than Shell\,1 (S1, I2, I3, S2, E2, EE2, and I4) in the plots of 6 \micron\, and 7.7 \micron \, features.
As seen in the top panel of the 6 \micron\, feature, the positions closer to the star show the peak wavelengths of $\sim 6.93-6.94$ \micron, which are close to 6.95-\micron\, C\,\textsc{iv} (18--16) and 6.98-\micron\, He\,\textsc{ii} (9--8) lines expected from the WC star.
Similarly, in the plot of the 7.7 \micron \, feature, positions close to the star show the peak wavelength of $\sim 7.46$ \micron, which is consistent with He\,\textsc{ii} (12--10) at 7.457 \micron. 
These lines show much smaller FWHM values (less than 0.15 \micron, corresponding to $\leq 6000$ km\,s$^{-1}$) compared to the positions farther than Shell\,1 (the middle panels of Figure \ref{fig:spefeature}).
We checked the line widths (FWHMs) to distinguish between the AIBs and atomic lines.
To investigate the line widths of the gas-phase atomic lines, we fitted the [S \textsc{iv}] line at 10.5 \micron\, with a Gaussian profile.
Its line widths are derived to be $\sim 0.16$ \micron \, and 0.12 \micron \, at S1 and S2, respectively.
Thus, the gas-phase atomic lines typically have FWHMs of $\sim0.1-0.15$ \micron, corresponding to $\sim 2860-4290$ km\,s$^{-1}$, around WR\,140. 
In summary, these lines close to the star position likely originate from the gas-phase atomic species, not AIBs.

The positions farther than Shell\,1 show the peak wavelengths at around 6.2--6.4 \micron\, and $\sim7.7$ \micron, and larger FWHMs (0.3 -- 0.6 \micron).
The I2 position shows a different peak wavelength from the other positions, which is caused by poor fittings due the spectral fringing artifact.
The 6 \micron \, and 7.7 \micron \, features farther than Shell\,1 could be attributed to AIBs.
The detected AIBs around WR\,140 show peak wavelengths close to 6.4 \micron.
The wavelength is obviously at a longer wavelength of the major AIB at 6.2\, \micron \, but close to one of the weak components seen in the Orion Bar spectra \citep{2024A&A...685A..75C, 2024A&A...685A..77P}.

The 7.7 \micron \, AIB consists of several components, and the main one is at 7.626 \micron \, in class A sources \citep{2024A&A...685A..75C}.
The AIBs around Shell\,1 and Shell\,2 show peak wavelengths around 7.73--7.74 \micron. 
The emission band shifts to longer wavelengths were observed in class B sources \citep{2002A&A...390.1089P}.

Both the 6 \micron \, and 7.7 \micron \, features around WR\,140 show peaks at longer wavelengths compared to those found in the Orion PDR or the ISM.
These shifts will be discussed in Section \ref{subsec:62feature}.

The 11.2 \micron \, AIB feature consists of two main components, 11.207 \micron \, and 11.25 \micron \, \citep{2024A&A...685A..75C, 2024A&A...685A..77P, 2025A&A...699A.133K}.
The observed 11.2 \micron \, spectra close to the star show their peaks at slightly longer wavelengths, and they are almost consistent with the 11.283-\micron\, He\,\textsc{i} (9-7) and 11.304-\micron\, He\,\textsc{ii} (18--14) features.
Contamination from He\,\textsc{i} and He\,\textsc{ii} lines, which are bright emission features from the core of WR\,140, may affect the 11.2 \micron\, AIB feature even at the positions farther than Shell\,1, possibly due to the extended diffraction pattern of the PSF. We also note that these He lines are commonly detected from WC stars \citep{2001AJ....121.2115S}.
As we mentioned before, the typical gas-phase atomic species show line widths around 0.1 -- 0.15 \micron, which are consistent with the observed line widths near the star position and S1.
The peak wavelength slightly shifts to a shorter wavelength compared to the He lines at S1, which may be affected by the outflow.
The He lines are likely more dominant at S1 compared to the farther positions.
By comparisons of peak wavelength and FWHM, the 11.2 \micron\, feature at positions around Shell\,2 (I3, S2, E2, EE2, and I4) likely originates from AIBs.
Note that the small differences in integrated intensity between on-shell and off-shell positions may reflect the possibility that the contribution from the foreground/background emission in the 11.2 \micron \, band, because we did not conduct background subtraction.

In summary, the 6 \micron \, and 7.7 \micron \, AIBs (the CC stretching modes) have been detected at positions at S1 on Shell\,1, the positions on Shell\,2, and off-shell positions farther than Shell\,1. 
Although there may be contamination from the He lines and/or from the foreground/background emission, the 11.2 \micron\, AIB (the CH out-of-plane bending mode) has also been seen around Shell\,2, including the off-shell positions (I2, I3, and I4).
The peak wavelengths of AIBs around WR\,140 are at longer wavelengths compared to class A emission bands. 
This suggests that the AIB features around WR\,140 are different from those observed in the general ISM.

The bottom panels of Figure \ref{fig:spefeature} show plots of the integrated intensities against projected distance from the central stars for three AIBs (6, 7.7, and 11.2 \micron). 
As discussed above, the 6 \micron, 7.7 \micron, and 11.2 \micron\, features are likely dominated by C and/or He lines at the positions of Star and in its vicinity (N1, E1, and I1).
In the three bottom panels of Figure \ref{fig:spefeature}, the star position (distance = 0 au) shows the largest value, and the integrated intensities decrease as the projected distance from the central stars increases.
The differences in integrated intensities between the core and each position are a factor of 17, 32, 20 for 6 \micron, 7, 12, 10 for 7.7 \micron, and 7, 17, 46 for 11.2 \micron \, features in order of I1 (903.6 au), N1 (1723.6 au), and E1 (1772.8 au), respectively.

As we discussed above, the two emission features at 6 \micron \, and 7.7 \micron \, are attributed to AIBs, which means that these emission features likely come from carbonaceous compounds at positions farther than Shell\,1.
The following two common patterns can be recognized in the plots of 6 \micron \, and 7.7 \micron \, features, both of which are attributed to the CC stretching modes:
(1) S1 on Shell\,1 is stronger than the three positions on Shell\,2, and
(2) These features show higher integrated intensities on Shell\,2 compared to off-shell positions (I3 and I4) when we compare the results among positions close to Shell\,2.

These patterns are not clear for the 11.2 \micron \, feature.
There are three possible explanations for the difference: (1) the 11.2 \micron \, feature is significantly dominated by atomic lines, (2) the 11.2 \micron\, feature originates from the CH out-of-plane bending mode, whereas the other two wavelength features are related to the CC stretching bond, and (3) the foreground/background emission significantly contributes.
We try to check the C-H vibration mode at 8.6 \micron, which is attributed to the CH in-plane bending mode.
However, as described before, the present data show a double-peak profile and cannot be studied reliably.  
Therefore, it is difficult at this stage to discuss the effect of the He lines on the 11.2 \micron \, band quantitatively.

\section{Discussion \label{subsec:dis}}

\subsection{Comparison to the ISM AIBs \label{subsec:class}}

\begin{figure*}[ht!]
\centering
\includegraphics[bb = 5 10 570 655, width = \textwidth]{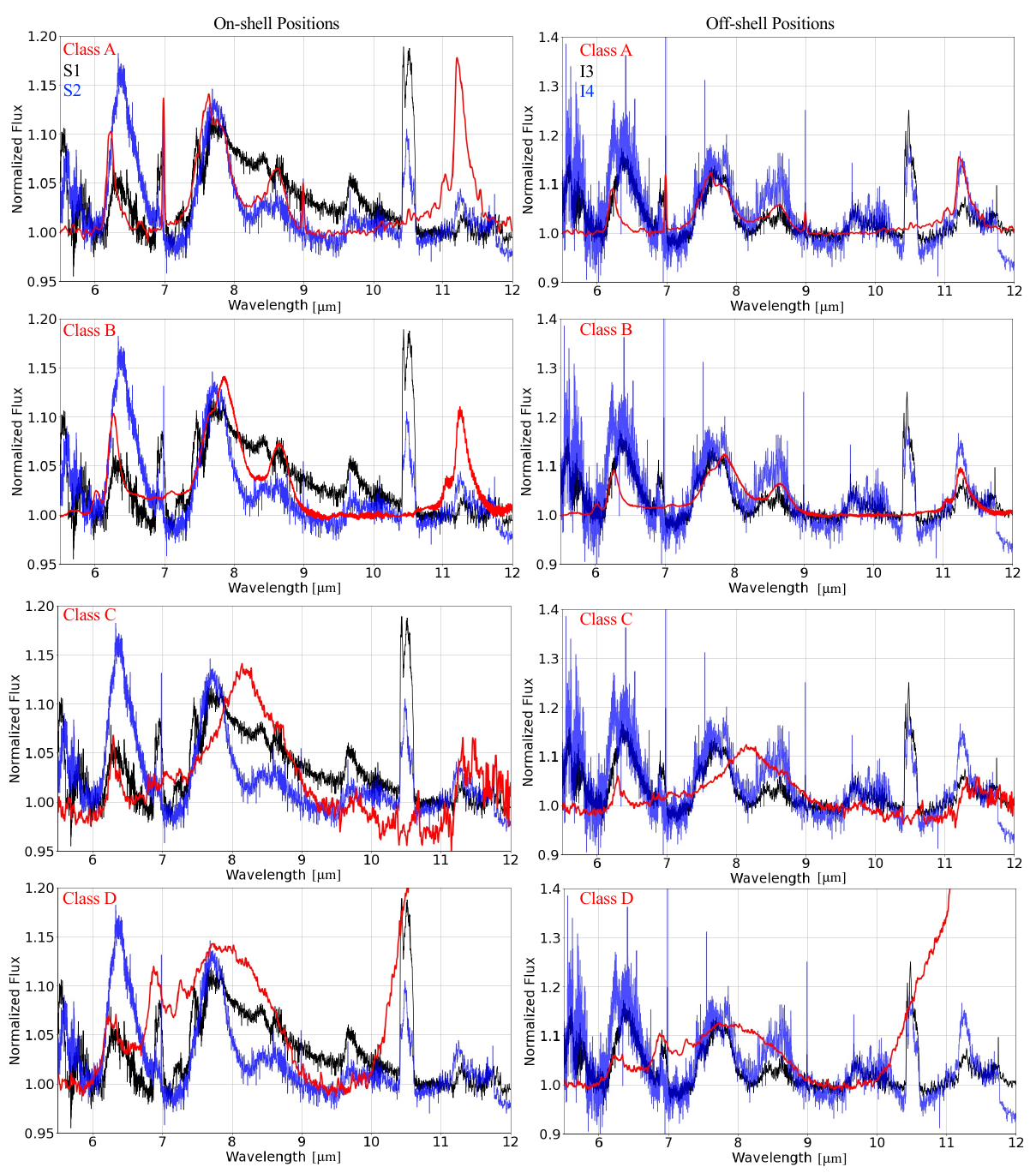}
\caption{Comparisons of the observed spectra (black and blue) and typical AIB emission features named classes A -- D (red). The left panels show observed spectra at two on-shell positions, S1 (black) and S2 (blue), and the right panels show those at two off-shell positions, I3 (black) and I4 (blue), respectively. The typical AIB emission features are as follows: class A is the compact \ion{H}{2} region IRAS\,23133+6050; class B is the post-AGB star HD\,44179; class C is the post-AGB star IRAS\,13416-6243 \citep{2002A&A...390.1089P}; class D is the post-AGB star IRAS\,22272+5435 \citep{2014MNRAS.439.1472M}.
\label{fig:class}}
\end{figure*}

The AIBs in the ISM are generally classified into four types.
On the other hand, the AIB class(es) associated with circumstellar dust around WR stars is still uncertain. 
There were some studies to try to investigate the AIB's features around WR stars \citep{2002ApJ...579L..91C, 2022ApJ...930..116E}.
\citet{2002ApJ...579L..91C} investigated the spectra of Infrared Space Observatory (ISO)/Short-Wavelength Spectrometer (SWS) spectra at 6.4 and 7.9 \micron\, toward WR\,48a, and they proposed that the 7.7 and 6.2 \micron\, features belong to classes B and C, respectively.
\citet{2022ApJ...930..116E} suggested that the 8 \micron\, features in the WC stars are likely related to class C.
However, these previous studies analyzed data with lower angular resolution where the emission from the bright core of the WC binary was blended with the circumstellar dust.
In this subsection, we compare the observed spectra around Shell\,1 and Shell\,2 of WR\,140 to the typical spectra of each class. 

Figure \ref{fig:class} shows comparisons of the observed spectra (black and blue) to the typical spectra of each class (red).
The left panels show spectra at two continuum peak positions on Shell\,1 and Shell\,2 (S1 and S2), and the right panels show those at off-shell positions (I3 and I4).
Here, we only show spectra at positions farther than Shell\,1, where the observed features are most likely to be attributed to the AIBs' emission rather than line emission (Section \ref{subsec:identify}).
We also exclude I2 due to the fringe patterns.
The observed 6 \micron\, features show longer peak wavelengths compared to classes A and B at all four positions.
The peak wavelengths of the observed 7.7 \micron\, features fall between classes A and B, and they are clearly different from class C at all of the positions. 

We will discuss these two features further in Section \ref{subsec:62feature}.

The 11.2 \micron \, emission is contaminated with the He lines, and it is difficult to reach a firm conclusion.
It appears that off-shell positions (I3 and I4) show clearer peaks.
Possible differences in the AIBs may imply that the CH bond increases at the off-shell positions.
We will investigate this possible difference further in Section \ref{subsec:spatial}.

\begin{figure}[ht!]
\centering
\includegraphics[bb = 0 10 380 450, scale = 0.6]{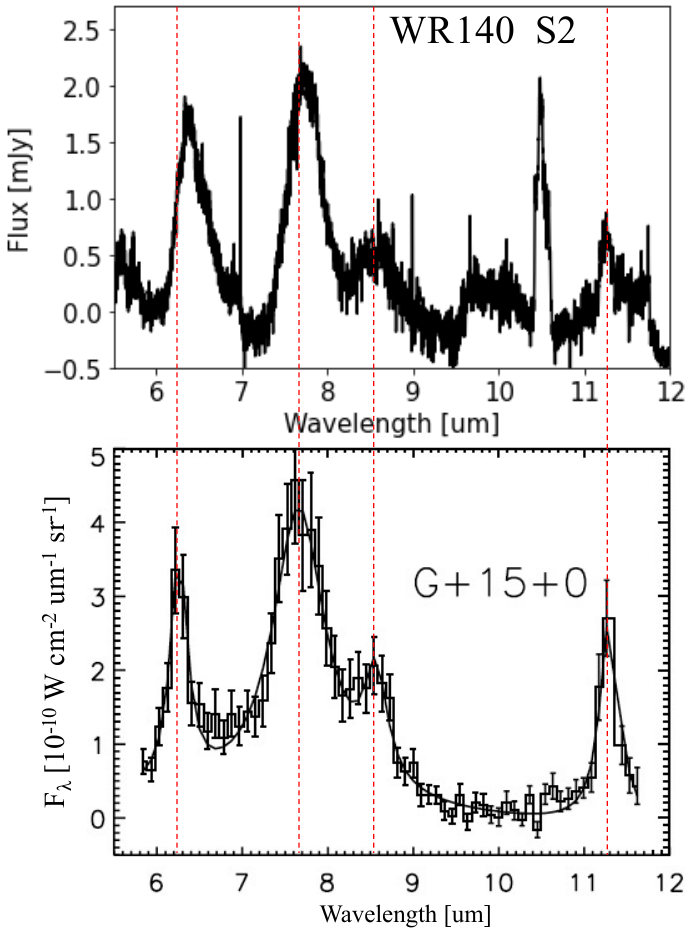}
\caption{Comparison of the spectra at WR\,140 S2 (top panel) and the diffuse ISM toward the Galactic plane obtained from \citet{2003A&A...405..999K} (bottom panel). The red dashed lines indicate the peak wavelengths attributed to AIBs. Note that there is a prominent [S IV] 10.51 $\mu$m emission line feature at S2 that is not associated to dust.
\label{fig:galactic_s2}}
\end{figure}

Figure \ref{fig:galactic_s2} shows a comparison of the spectra between S2 and the diffuse ISM toward the Galactic plane, which is obtained from Figure 3b of \citet{2003A&A...405..999K}.
The S2 spectral features resemble those of the diffuse ISM in our Galaxy, but the peak wavelengths of the 6 \micron \, and 7.7 \micron \, features in WR\,140 are found at slightly longer wavelengths compared to the diffuse ISM.
The spectrum at S2 seems to imply a link between the dust produced in WR 140 to the diffuse ISM, and that further processing of the WR\,140 dust in the ISM causes a shift to shorter wavelengths corresponding to class A.

\subsection{Spatial Variation of Integrated Intensities of PAH Emission \label{subsec:spatial}}

\begin{figure*}[ht!]
\includegraphics[bb = 5 10 560 130, width=\textwidth]{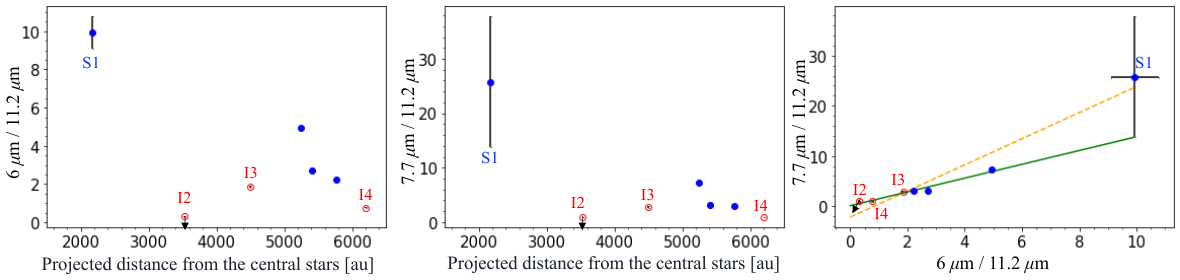}
\caption{Plots using integrated-intensity ratios. The left and middle panels show the dependence of the 6 \micron/11.2 \micron\, and 7.7 \micron/11.2 \micron\, integrated-intensity ratios on the projected distance from the central stars, respectively. The blue-filled and red-open circles represent on-shell and off-shell positions, respectively. The points without their position names indicate the positions on Shell\,2 (S2, EE2, E2). The right panel indicates a positive correlation between these two integrated-intensity ratios. The orange dashed line indicates the best-fit function including S1 ($y = 2.6004x - 2.1795$), whereas the green line shows the best-fit function excluding S1 ($y = 1.374x + 0.0836$). The Pearson's correlation coefficients $r$ for these pairs are derived to be 0.972 and 0.985 for with S1 and without S1, respectively.  
\label{fig:integ}}
\end{figure*}

The left and middle panels of Figure \ref{fig:integ} show the distance from the star vs. the integrated-intensity ratios of the 6 \micron/11.2 \micron\, and 7.7 \micron/11.2 \micron\, respectively.
As we noted before, the I2 position shows fringes around 5--8 \micron, and the 6 \micron\, and 7.7 \micron\, features have not been clearly detected. 
Hence, its plots are upper limits.
The S1 shows the highest integrated-intensity ratios.
This could be attributed to the possibility that the 11.2 \micron \, feature at S1 is significantly dominated by the He lines rather than the AIB (Section \ref{subsec:identify}); a narrow FWHM of the 11.2 \micron \, feature at S1 leads to large integrated-intensity ratios.

The left and middle panels of Figure \ref{fig:integ} show that the off-shell positions (I2, I3, and I4) show relatively lower values than those of the on-shell positions.
This suggests that the 11.2 \micron\, CH out-of-plane bending mode becomes stronger at the off-shell positions compared to the CC stretching modes (6 \micron\, and 7.7 \micron).
As we noted before, the 11.2 \micron \, feature may be affected by the foreground/background emission, and it is not easy to quantitatively discuss. 
However, if it is true that the 11.2 \micron \, feature would be stronger at the off-shell positions, one possible scenario for how the structures of carbonaceous compounds change is that the hydrogen-rich stellar wind from the O-star companion interacts with the carbon-rich dust shells, and the number of C-H bonds increases.  
The nature of the 11.2 \micron\, feature around WR\,140 will be investigated in further detail by Senoo et al.~(in prep.).

\subsection{Correlations between AIBs \label{subsec:correlation}}

It is known that the integrated intensities of 6 \micron\, and 7.7 \micron\, AIBs normalized by the 11.2 \micron\, AIB show positive correlations \citep{2008ARA&A..46..289T}.
The right panel of Figure \ref{fig:integ} shows the relationship between the integrated-intensity ratios of 6 \micron/11.2 \micron\, and 7.7 \micron/11.2 \micron. 

We derived the Pearson's correlation coefficient $r$ for these pairs.
They are 0.972, including S1, increasing to 0.985, excluding S1.
These results indicate a strong positive correlation between them.
Since the 11.2 \micron\, feature at S1 seems to be dominated by the He lines, the fitting result including S1 likely becomes worse compared to that excluding S1.

In summary, the positive correlation between 6 \micron\, and 7.7 \micron\, is seen in the harsh environment around WR\,140.
This correlation seems to support that these two bands trace AIBs.
On the other hand, this result does not prove that the 11.2 \micron \, feature is attributed to AIBs. 
As we mentioned earlier, the 11.2 \micron \, feature may be contaminated with the He lines. 
Then, it is difficult to compare the absolute values of the intensity of the 11.2 \micron \, band.

\subsection{Relationship between FWHM and Peak Wavelength of 6 \micron\, and 7.7 \micron\, Features  \label{subsec:62feature}}

\begin{figure*}[ht!]
\centering
\includegraphics[bb = 0 10 530 400, scale = 0.6]{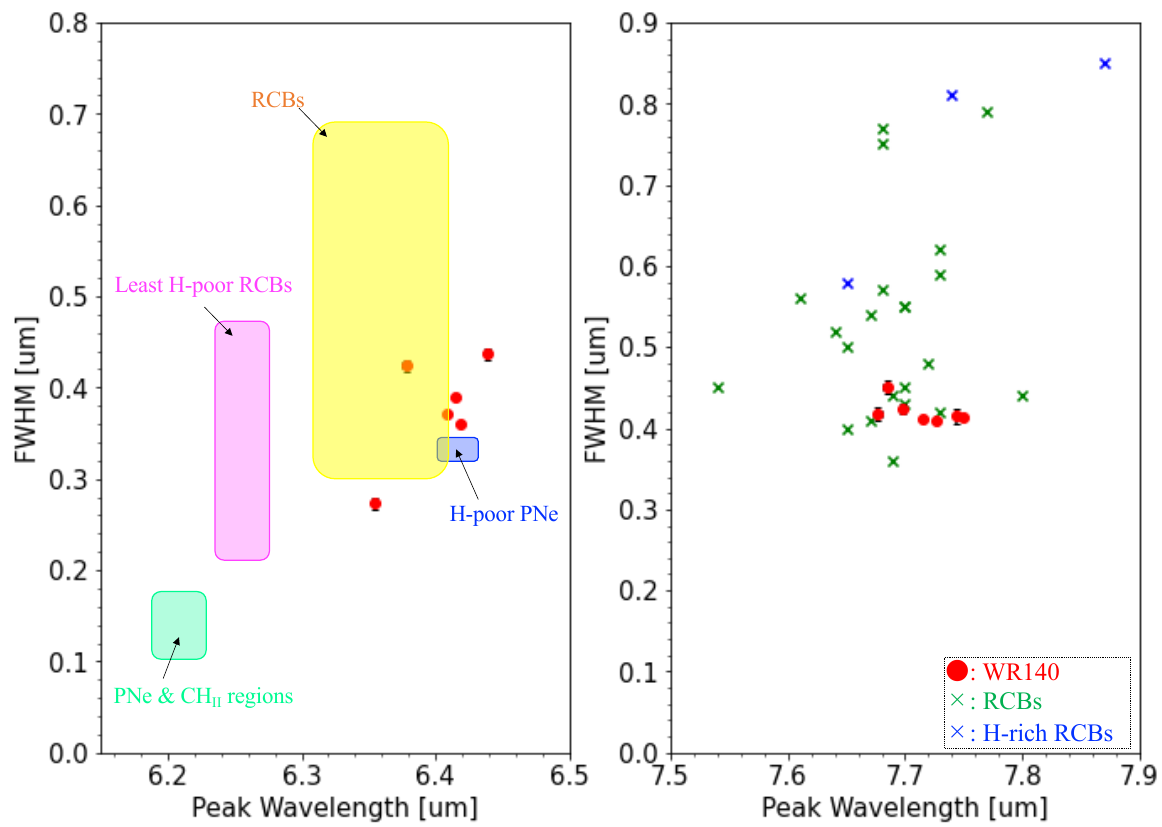}
\caption{Relationship between the peak wavelength and FWHM for the 6 \micron\, and 7.7 \micron\, features. (Left) The plot for the 6 \micron\, feature. Red data points are our results around WR\,140. The other shadow regions are taken from \citet{2013ApJ...773..107G}. (Right) The plot for the 7.7 \micron\, feature. The red circles indicate our results around WR\,140, and the other cross points are taken from Tables 2 and 3 of \citet{2013ApJ...773..107G}, respectively.
\label{fig:62feature}}
\end{figure*}

\citet{2013ApJ...773..107G} observed 31 R Coronae Borealis (RCB) stars with {\it {Spitzer}}.
Hydrogen is extremely deficient in most RCBs, and these sources show an unpredictable optical variability with relatively rapid declines in brightness.
\citet{2013ApJ...773..107G} found clear differences in the peak wavelength of the 6 \micron\, feature and its FWHM among source types; RCBs show larger FWHMs ($\sim0.35-0.7$ \micron) and longer peak wavelengths ($\sim 6.32-6.4$ \micron) compared to the other type sources such as planetary nebulae (PNe) and compact \ion{H}{2} regions \citep[see Figure 15 in][]{2013ApJ...773..107G}.
Reasons why hydrogen-poor sources show longer peaks and wider line widths are not yet understood.
In this subsection, we compare the peak wavelengths and FWHMs of the 6 \micron\, and 7.7 \micron\, features around WR\,140 to the results presented by \citet{2013ApJ...773..107G} and examine a possible link to the chemical environment of WR\,140.

The left panel of Figure \ref{fig:62feature} shows the relationship between the peak wavelength and FWHM of the 6 \micron\, feature.
The red points show our results around WR\,140 and the color-shadowed regions indicate the ranges for each source type taken from \citet{2013ApJ...773..107G}.
Here, we exclude I2 data due to large uncertainties.
The WR\,140 data points are located at the lower right corner of the range of the RCBs.

The right panel of Figure \ref{fig:62feature} shows the same plot but for the 7.7 \micron\, feature of WR\,140 and RCBs.
The values of RCBs (crosses) are taken from Tables 2 and 3 in \citet{2013ApJ...773..107G}\footnote[2]{They categorized RCBs into two groups based on the interstellar reddening $E(B-V)$; low ($ \leq 0.3$) and high ($ > 0.3$) cases, respectively. Here we treat both cases as RCBs without further classification.}.
The results around WR\,140 agree with the lower part of the RCBs (green crosses).
On the other hand, WR\,140 shows smaller FWHMs than those in hydrogen (H)-rich RCBs (blue crosses).

To check whether these tendencies are real, we compare the average values.
The average peak wavelengths (and their uncertainties) are 7.69 (0.04), 7.75 (0.06), and 7.71 (0.02) \micron\, for RCBs, H-rich RCBs, and WR\,140, respectively.
The average FWHMs are 0.53 (0.09), 0.75 (0.11), and 0.420 (0.009) \micron\, for the same order. 
Although the sample size of the H-rich RCBs is small, the tendencies are likely real.
In short, the peak wavelength is not clear, whereas FWHMs around WR\,140 are marginally consistent with the H-poor RCBs but differ from the H-rich RCBs.

From the comparisons of spectral features of the 6 \micron\, and 7.7 \micron\, bands among WR\,140 and RCBs, we found that both features in WR\,140 are similar to those of RCBs with H-poor conditions.
This is a reasonable result since C-rich WR (WC) stars typically have hydrogen-poor conditions (Section \ref{sec:intro}).

\subsection{Chemistry of Carbonaceous Compounds around WR\,140 \label{subsec:PAHWR140}}

We have shown that the AIBs at the 6 \micron\, and 7.7 \micron\, have been detected at positions farther than Shell\,1. 
The observing date (2022 July 8 UT) corresponds to $\sim5.5$ years after the last dust-formation episode and periastron passage on 2016 December, creating Shell\,1 \citep{2022NatAs...6.1308L, 2021MNRAS.504.5221T}. 
Since the orbital period is 7.93 yr, Shell\,2 was formed $\sim13.43$ years ago at the observing timing.
The presence of the AIBs around Shell\,1 suggests that the formation of carbonaceous compounds occurs in a very short timescale ($\leq5$ years).
The growth of dust grains through accretion from the gas phase on this timescale indicates that the gas density is many orders of magnitude higher than that in the ISM \citep{2012MNRAS.422.1263H}.
These regions have an extreme environment ($i.e.,$ very hot gas) because He\,\textsc{i} or He\,\textsc{ii} lines have been detected there.
These physical conditions are significantly different from those of the ISM and regions where previous studies focused \citep{2008ARA&A..46..289T, 2013A&A...558A..62J}. 
Consequently, the other kind of chemistry of carbonaceous compounds likely forms around WR\,140.

We compared the observed spectra to the typical spectra named classes A--D (see Section \ref{subsec:class}).
The observed spectra around WR\,140 do not coincide with the spectral features of these typical classes. 
The features of AIBs, $i.e.,$ peak wavelength and width, change depending on the structures of carbonaceous compounds \citep{2005ApJ...620L.135D, 2012ApJ...761..115D, 2012A&A...548A..40C}.
Hence, the spectral features observed around WR\,140 suggest that the structures of carbonaceous compounds around WR\,140 are different from other types of astronomical sources.

When we focus on FWHM and peak wavelengths of the 6 \micron\, and 7.7 \micron\, features, we found similarities between WR\,140 and hydrogen-poor RCBs (Section \ref{subsec:62feature}).
These similarities suggest that the carbonaceous compounds around WR\,140 are linked to those formed in the hydrogen-poor conditions. 
This is consistent with the expected conditions of typical WC stars.

Hydrogen could be provided by the companion O star and is not completely absent around WR\,140. 
Although we believe the 11.2 \micron\, AIB is identified around Shell 2, it could be contaminated by \ion{He}{1} and/or \ion{He}{2} lines.

In summary, the results imply that hydrogen-poor carbonaceous compounds are initially formed within a short time scale (less than $\sim5$ years). 
The impact of the hydrogen-rich stellar wind from the O star provides hydrogen to the carbonaceous compounds, and these hydrogen-attached carbonaceous compounds survive at more distant positions from the WR star.
These carbonaceous compounds may be the origin of the interstellar PAHs by further processing by massive stars or in the diffuse ISM.

\section{Conclusions \label{sec:con}}

We present spectral data obtained with JWST/MRS toward the dust-forming WC star WR\,140.
The data cover the latest and the second dust shells (Shell\,1 and Shell\,2, respectively).
The main conclusions of this paper are summarized below.

\begin{enumerate}
\item The 6 \micron\, and 7.7 \micron\, AIBs, which are attributed to the C-C stretching mode, have been detected at positions farther than Shell\,1. These features are particularly prominent in the on-shell positions compared to the off-shell positions.

\item The 11.2 \micron\, AIB, the C-H out-of-plane bending mode, is tentatively detected at the positions around Shell\,2, although they are possibly contaminated with the He lines. The line width of this feature at S1 agrees with the atomic gas-phase species, and we could not identify the 11.2 \micron \, AIB on Shell\,1.

\item The spectra around WR\,140 do not coincide with typical spectra of classes A--D. These results imply that the structures of carbonaceous compounds around WR\,140 are different from those in the other astronomical objects.

\item The integrated intensities of 6 \micron\, and 7.7 \micron\, AIBs normalized by 11.2 \micron\, show a strong positive correlation ($r\approx0.98$), which can be seen in the other astronomical objects.

\item We found that the peak wavelengths and FWHMs of the 6 \micron\, and 7.7 \micron\, AIBs around WR\,140 agree with those around hydrogen (H)-poor RCBs. These results suggest that carbonaceous compounds around WR\,140 are mainly formed in the H-poor conditions without significant contamination by the H-rich colliding wind of the O-star companion.
This could be realized around the WC star.
\end{enumerate}

From these observational results, we propose that the possible formation mechanisms of carbonaceous compounds around WR\,140.
Hydrogen-poor carbonaceous compounds are formed initially within a short timescale ($\leq 5$ yr) after the wind-wind collision.
They are the origin of the 6 \micron \, and 7.7 \micron \, AIBs.
Hydrogen is provided by the stellar wind from the companion O star, and the hydrogen-attached carbonaceous compounds are formed and survive around Shell\,2, and maybe farther out than this position.
The 11.2 \micron \, AIB would come from these materials.
These carbonaceous compounds may be the origin of the interstellar AIB after further processing in the ISM.

\begin{acknowledgments}
K.T. is supported by JSPS KAKENHI grant Nos. 21H01142, 24K17096, and 24H00252.  
T.O. acknowledges the support by JSPS KAKENHI grant No. JP24K07087.
N.D.R. is grateful for support from the Cottrell Scholar Award \#CS-CSA-2023-143 sponsored by the Research Corporation for Science Advancement and from Space Telescope Science Institute through JWST Guest Observer program \#4093.
R.S. is supported by JSPS KAKENHI grant No. 25KJ0830.
J.S.-B. acknowledges the support received by the UNAM DGAPA-PAPIIT project AG 101025.
We thank Dr. Sergey V. Marchenko and Dr. Greg Sloan for giving their thoughtful comments on the draft. We also thank the anonymous referee for their valuable feedback that has improve the quality of this work.

This work is based on observations made with the NASA/ESA/CSA James Webb Space Telescope. The data were obtained from the Mikulski Archive for Space Telescopes at the Space Telescope Science Institute, which is operated by the Association of Universities for Research in Astronomy, Inc., under NASA contract NAS 5-03127 for JWST. These observations are associated with program \#1349.
Support for program \#1349 was provided by NASA through a grant from the Space Telescope Science Institute, which is operated by the Association of Universities for Research in Astronomy, Inc., under NASA contract NAS 5-03127.
The material is based upon work supported by NASA under award number 80GSFC21M0002.
\end{acknowledgments}

%

\vspace{5mm}
\facilities{James Webb Space Telescope (JWST)}





\appendix

\section{Continuum subtraction and Gaussian fitting} \label{appA}

\begin{figure*}[ht!]
\includegraphics[bb = 10 0 580 620, width=\textwidth]{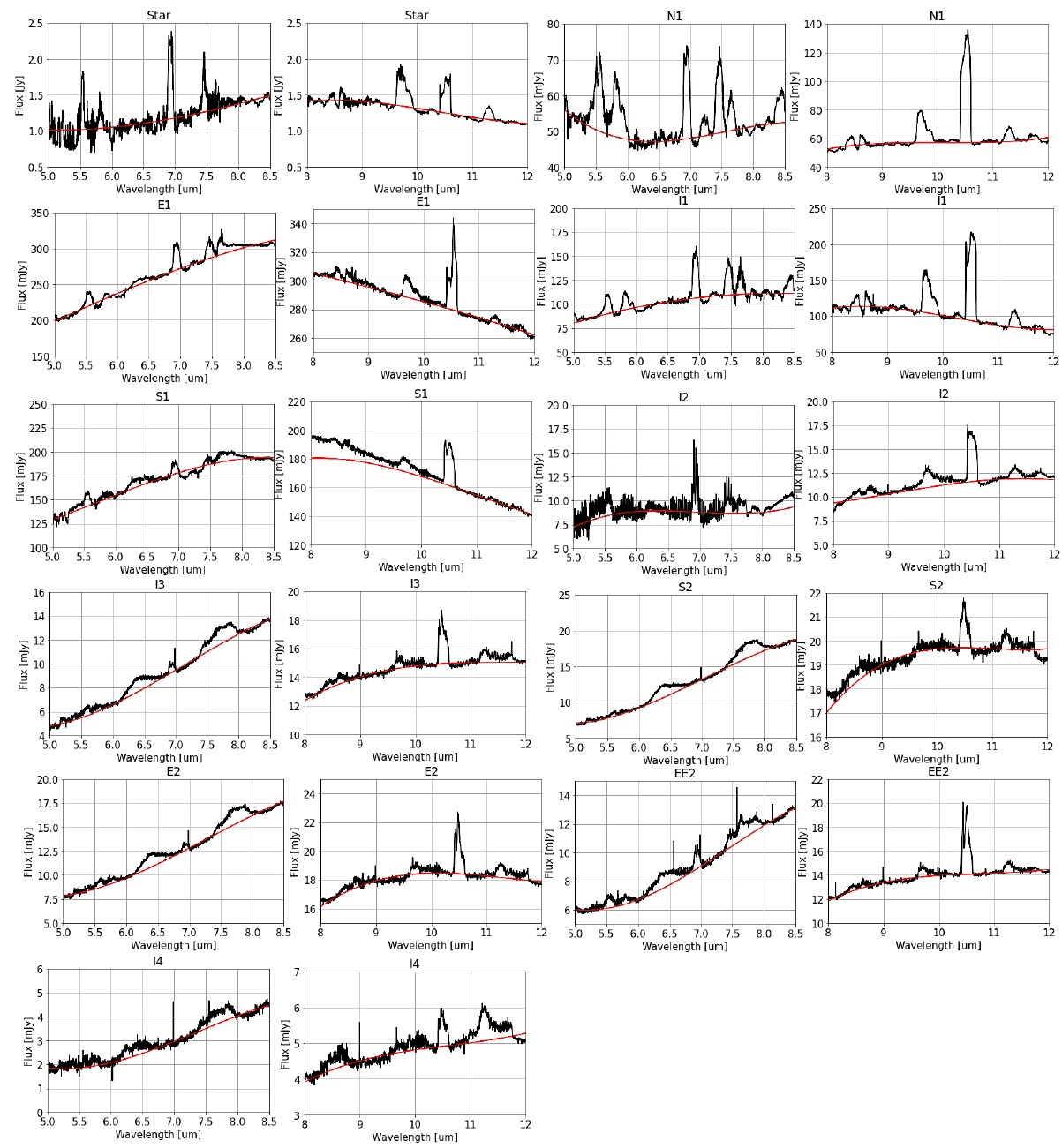}
\caption{Continuum fitting for the observed spectra. Black and red lines indicate the observed spectra and continuum fitting, respectively. The fitting results for the wavelength range of 8.0--12.0 \micron\ are used for analysis only of the 11.3 \micron\ feature.
\label{fig:contfit}}
\end{figure*}

We applied a spline function for continuum subtraction.
We subtracted the continuum emission dividing wavelengths into different ranges (5--8.5 \micron\, and 8--12 \micron), except for the simultaneous fitting for 7.7 \micron\, and 8.6 \micron\, at S1 and S2.
Figure \ref{fig:contfit} shows continuum fitting (red lines) for the observed spectra (black lines).

\begin{figure*}[ht!]
\includegraphics[bb = 0 10 570 200, width=\textwidth]{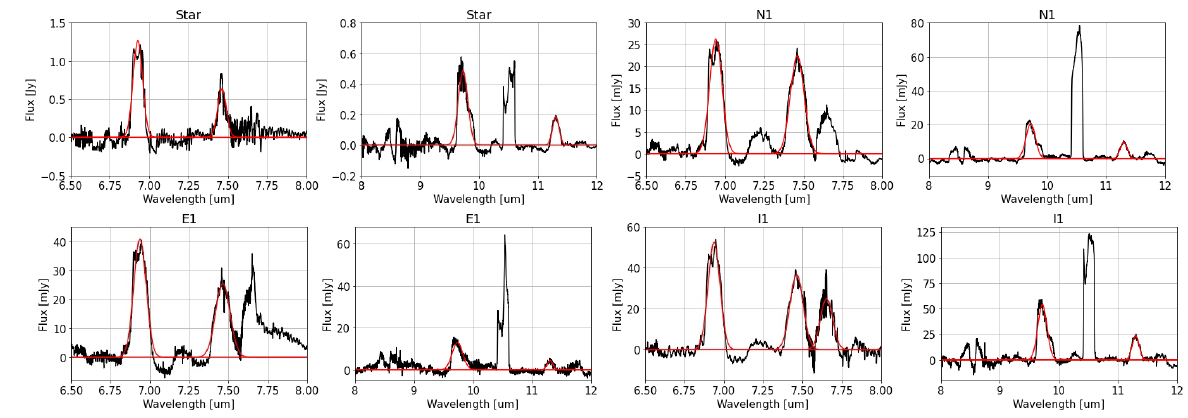}
\caption{Gaussian fitting results toward positions at the star WR\,140 and close to it (N1, E1, and I1). The AIBs have not been identified at these positions.
\label{fig:Gauss1}}
\end{figure*}

\begin{figure*}[ht!]
\includegraphics[bb = 0 10 570 600, width=\textwidth]{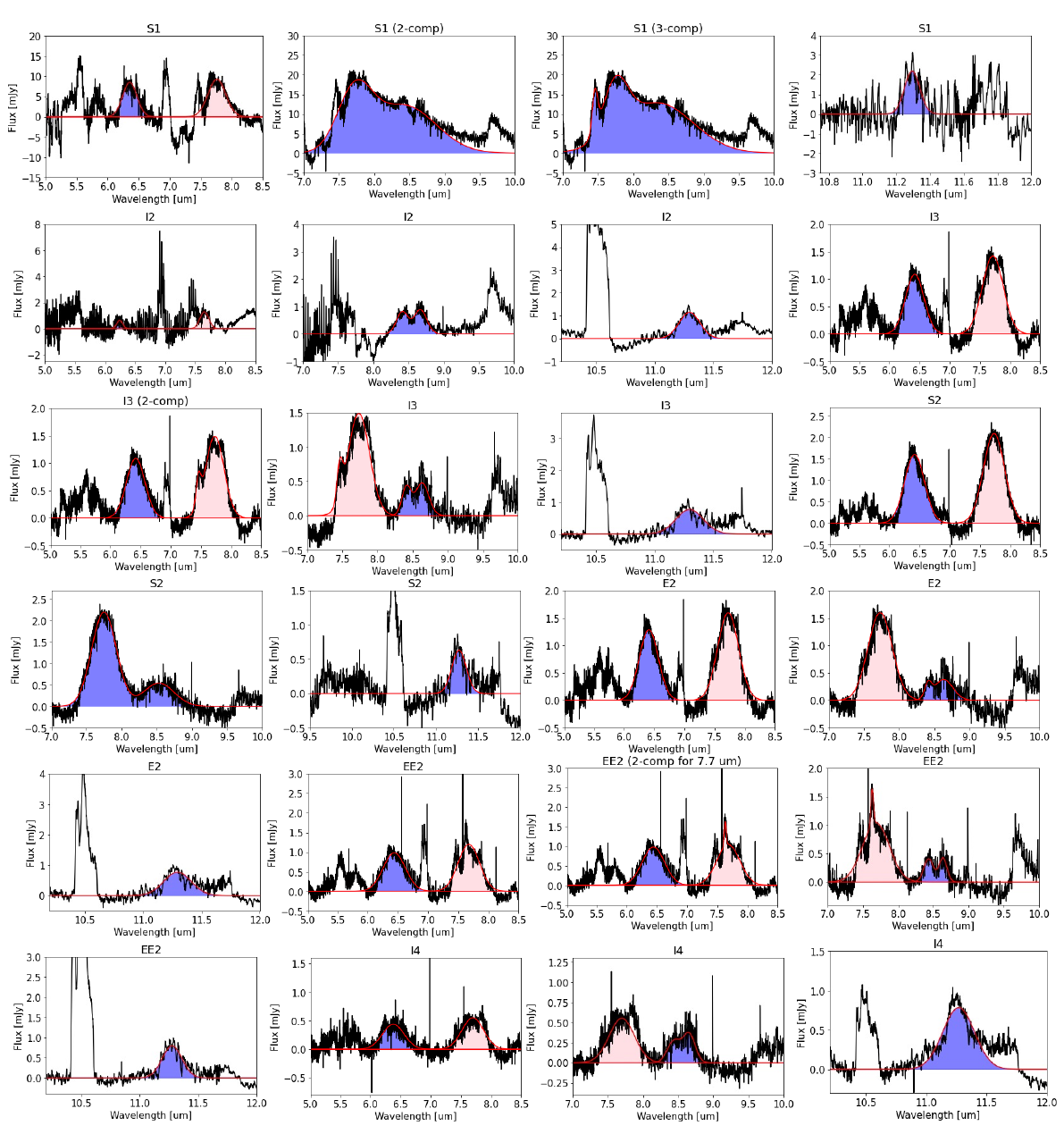}
\caption{Gaussian fitting results toward positions where the AIBs have been detected. Different filled colors mean that they are treated independently.
\label{fig:Gauss2}}
\end{figure*}

After continuum subtraction, we fitted spectra with a Gaussian profile.
Table \ref{tab:gaus} summarizes the obtained line parameters.
Figures \ref{fig:Gauss1} and \ref{fig:Gauss2} show the fitting results.
Figure \ref{fig:Gauss1} shows spectra at WR\,140 (labeled as Star) and positions close to WR\,140 (within 1800 au). 
We could not identify any AIBs at these positions.

Figure \ref{fig:Gauss2} shows spectra at positions farther than Shell\,1. 
At the S1 position, we try to perform multi-Gaussian fitting (2 and 3 cases) for 7.7--8.6 \micron \, and found that the 3-component fitting is the best result. Then, we utilize the results of the 3-component fitting in this paper.
All of the spectra show the double-peak features at 8.6 \micron, which prevents us from firmly identifying the AIB. 

\startlongtable
\begin{deluxetable}{lcccc}
\tablewidth{0pt} 
\tablecaption{Line Parameters obtained from the Gaussian fitting \label{tab:gaus}}
\tablehead{
\colhead{Position} & \colhead{Peak Intensity } & \colhead{Peak Wavelength} & \colhead{FWHM} & \colhead{Integrated Intensity} \\
\colhead{} & \colhead{($\mu$Jy)} & \colhead{(\micron)} & \colhead{(\micron)} & \colhead{($\mu$Jy\, \micron)}
}
\startdata 
\multicolumn{5}{c}{\bf {6 \micron\, feature}} \\
Star & $1171767 \pm 26020$	& $6.9278 \pm 0.0007$ & $0.0660 \pm 0.0017$ & $82293 \pm 2791$ \\
N1	& $26326 \pm 351$ &	$6.9416 \pm	0.0006$	& $0.0920 \pm 0.0014$ & $2578 \pm 52$ \\
E1	& $40895 \pm 374$ & $6.9373	\pm 0.0004$ & $0.0925 \pm 0.0010$ & $4028 \pm 56$ \\
I1	& $52569 \pm 680$ &	$6.9377 \pm	0.0006$ & $0.0888 \pm 0.0013$ & $4969 \pm 98$ \\
S1  & $8551 \pm	141$ & $6.354 \pm 0.002$ & $0.273 \pm 0.007$ & $2483 \pm 74$ \\		
I2  & $631 \pm 43$ & $6.241 \pm 0.004$ & $0.117 \pm 0.009$ & $78 \pm 8$ \\							
I3	& $1090	\pm 7$ & $6.418	\pm 0.001$ & $0.361 \pm	0.003$ & $418 \pm 4$ \\					
S2	& $1601 \pm 9$ & $6.415 \pm	0.001$ & $0.390 \pm	0.003$ & $665 \pm 6$ \\							
E2  & $1287 \pm	9$ & $6.409 \pm 0.001$ & $0.371 \pm 0.003$ & $509 \pm 5$ \\						
EE2	& $985 \pm 11$ & $6.439 \pm	0.002$ & $0.437	\pm 0.006$ & $458 \pm 8$ \\
I4	& $440 \pm 6$ & $6.378 \pm 0.003$ & $0.424 \pm 0.007$ &	$198 \pm 4$ \\
\multicolumn{5}{c}{\bf {7.7 \micron\, feature}} \\
Star & $598790 \pm 18689$ &	$7.4604 \pm 0.0007$ & $0.045 \pm 0.002$ & $28479 \pm 1356$ \\
N1 & $22460	\pm 239$ & $7.4598 \pm 0.0005$ & $0.103 \pm 0.001$ & $2466 \pm 38$ \\
E1 & $25439 \pm 535$ & $7.4642 \pm 0.0011$ & $0.108 \pm 0.002$ & $2935 \pm 90$ \\
I1 & $36726 \pm	473$ & $7.4626 \pm 0.0006$ & $0.099 \pm	0.001$ & $ 3852 \pm 72$ \\							
S1 & $9316 \pm 158$ & $7.761 \pm 0.003$ & $0.3381 \pm 0.0069$ & $3353 \pm 89$ \\
S1 (2-comp)\tablenotemark{a} & $14961 \pm 254$ & $7.7202 \pm 0.003$ & $0.5746 \pm 0.0118$ & $9151 \pm 244$ \\
S1 (3-comp)\tablenotemark{b} & $13216 \pm 224$ & $7.7445 \pm 0.003$ & $0.4149 \pm 0.0085$ & $5837 \pm 155$ \\
S1 (3-comp)\tablenotemark{b} & $10252 \pm 174$ & $7.4542 \pm	0.002$ & $0.0906 \pm 0.0019$ & $989 \pm 26$ \\			
I2 & $1291 \pm 31$ & $7.649 \pm	0.002$ & $0.17 \pm 0.01$ & $239 \pm 10$ \\							
I3 & $1423 \pm 12$ & $7.715 \pm 0.002$ & $0.412 \pm 0.004$ & $624 \pm 8$ \\
I3 (2-comp) & $512 \pm 4$ & $7.460 \pm 0.002$ & $0.094 \pm	0.001$ & $51 \pm 1$ \\
I3 (2-comp)	& $1491	\pm 13$ & $7.731 \pm 0.002$ & $0.364 \pm 0.004$ & $578 \pm 8$ \\
S2 & $2200 \pm 11$ & $7.749 \pm	0.001$ & $0.414 \pm 0.002$ & $971 \pm 7$ \\						
E2 & $1597 \pm	11$ & $7.7261 \pm 0.0013$ & $0.4091 \pm 0.0033$ & $696 \pm 7$ \\		
EE2 &$1202 \pm 17$ & $7.677 \pm 0.003$ & $0.4185 \pm 0.0076$ & $536 \pm 12$ \\
EE2 (2-comp) & $631	\pm 9$ & $7.632 \pm 0.003$ &  $0.0419 \pm 0.0008$ & $28 \pm	1$ \\
EE2 (2-comp) & $1045 \pm	15$ & $7.685 \pm 0.003$ & $0.450 \pm 0.008$ & $500 \pm 11$ \\
I4	& $553\pm 7$ & $7.699 \pm 0.003$ & $0.4238 \pm 0.0068$ & $249 \pm 5$ \\
\multicolumn{5}{c}{\bf {8.6 \micron\, feature}} \\
S1 (2-comp)\tablenotemark{a} & $12149 \pm 675$ &	$8.429 \pm 0.009$ & $1.06 \pm 0.09$ & $13664 \pm 1340$ \\
S1 (3-comp)\tablenotemark{b} & $12838 \pm 713$ & $8.342 \pm 0.009$ & $1.2 \pm 0.1$ &	$16324 \pm 1601$ \\
I2 & $770 \pm 18$ & $8.574 \pm 0.012$ & $0.64 \pm 0.07$ & $528 \pm 58$ \\
I2 (2-comp)	& $807 \pm 45$ & $8.411 \pm 0.009$ & $0.195 \pm 0.016$ & $167 \pm 16$ \\
I2 (2-comp) & $880 \pm 49$ & $8.669 \pm	0.009$ & $0.214 \pm	0.017$ & $200 \pm 20$ \\
I3 & $483 \pm 13$ &	$8.624 \pm 0.004$ &	$0.245 \pm 0.011$ &	$126 \pm 7$ \\
I3 (2-comp)	& $410 \pm 34$ & $8.410 \pm	0.009$ & $0.155 \pm 0.019$ & $67 \pm 10$ \\
I3 (2-comp)	& $478 \pm 23$ & $8.633 \pm	0.010$ & $0.225 \pm	0.021$ & $115 \pm 12$ \\
S2 & $542 \pm 10$ &	$8.529 \pm 0.005$ & $0.508 \pm 0.012$ &	$293 \pm 9$ \\
E2 & $490 \pm 20$ &	$8.634 \pm 0.004$ & $0.179 \pm 0.014$ &	$93 \pm 8$ \\
E2 (2-comp)	& $260 \pm 14$ & $8.409 \pm	0.009$ & $0.12 \pm 0.01$ & $33 \pm 3$ \\
E2 (2-comp) & $378 \pm 21$ & $8.631 \pm 0.009$ & $0.313 \pm	0.025$ & $126 \pm 12$ \\	
EE2 (2-comp) & $412	\pm 18$ & $8.433 \pm 0.004$ & $0.161 \pm 0.012$ & $70 \pm 6$ \\
EE2 (2-comp) & $422 \pm 18$ & $8.637 \pm 0.004$ & $0.134 \pm 0.007$ & $60 \pm 4$ \\
I4 (2-comp)	& $271 \pm 18$ & $8.44 \pm 0.02$ & $0.22 \pm 0.03$ & $64 \pm 9$ \\
I4 (2-comp) & $353 \pm 24$ & $8.656 \pm 0.011$ & $0.203 \pm	0.016$ & $76 \pm 8$ \\
\multicolumn{5}{c}{\bf {11.2 \micron\, feature}} \\
Star & $187484 \pm 3557$ & $11.305 \pm 0.001$ & $0.127 \pm 0.003$ & $25264 \pm 732$ \\
N1 & $9808 \pm 81$ & $11.299 \pm 0.001$ & $0.140 \pm 0.001$ & $1464 \pm 19$ \\
E1 & $4558 \pm 157$ & $11.304 \pm 0.002$ & $0.114 \pm 0.005$ & $551 \pm 29$ \\
I1 & $23705 \pm 245$ & $11.303 \pm 0.001$ &	$0.138 \pm 0.002$ & $3477 \pm 55$ \\							
S1 & $2189 \pm 112$ & $11.294 \pm 0.003$ & $0.107 \pm 0.006$ & $250 \pm 20$ \\
I2 & $1142 \pm 16$ & $11.295 \pm 0.001$ & $0.209 \pm 0.004$ & $254 \pm 6$ \\
I3 & $768 \pm 17$ &	$11.301 \pm	0.003$ & $0.275 \pm	0.008$ & $225 \pm 8$ \\
S2 & $631 \pm 17$ & $11.271 \pm	0.003$ & $0.200 \pm	0.006$ & $134 \pm 6$ \\
E2 & $772 \pm 17$ &	$11.296 \pm	0.003$ & $0.278 \pm	0.008$ & $229 \pm 8$ \\
EE2	& $796 \pm 15$ & $11.279 \pm 0.002$ & $0.200 \pm 0.004$ & $170 \pm 5$ \\
I4 & $785 \pm 11$ &	$11.270 \pm	0.002$ & $0.311 \pm	0.006$ & $260 \pm 6$ \\
\enddata
\tablecomments{Errors indicate the standard deviation derived from the Gaussian fitting. The note of (2-comp) indicates that two-component Gaussian fitting derives the line parameters for each adjacent pair.}
\tablenotetext{a}{These two components are fitted simultaneously.}
\tablenotetext{b}{These three components are fitted simultaneously.} 
\end{deluxetable}

\bibliography{WR140_PAH}{}
\bibliographystyle{aasjournal}



\end{document}